\documentclass[10pt]{sigplanconf}

\usepackage{amsmath}

\usepackage{graphicx}
\usepackage{mathtools}
\usepackage{tabularx}
\usepackage{url}
\usepackage{color}

\newcommand{\D}{\mathbf{D}}
\newcommand{\y}{\mathbf{y}}
\newcommand{\x}{\mathbf{x}}
\newcommand{\dt}{\mathit{dt}}
\newcommand{\DT}{\mathit{DT}}
\newcommand{\CH}{\mathit{H}}

\newcommand{\CV}[1]{\mathit{CV^{#1}}}
\newcommand{\OM}[1]{\mathit{OM^{#1}}}
\newcommand{\varst}{\mathit{varstatus}}

\newcommand{\prov}{\mathit{prov}}
\newcommand{\diffin}{\mathit{diff}_{in}}
\newcommand{\diffd}{\mathit{diff}_d}
\newcommand{\diffo}{\mathit{diff}_{out}}
\newcommand{\diffOM}{\mathit{diff}_{OM}}
\newcommand{\diffCV}{\mathit{diff}_{CV}}
\newcommand{\recomp}{\textit{ReComp}}
\newcommand{\genes}{\mathit{genes}\ }

\begin{document}

\setlength{\pdfpageheight}{\paperheight}
\setlength{\pdfpagewidth}{\paperwidth}

\conferenceinfo{CONF 'yy}{Month d--d, 20yy, City, ST, Country}
\copyrightyear{20yy}
\copyrightdata{978-1-nnnn-nnnn-n/yy/mm}
\copyrightdoi{nnnnnnn.nnnnnnn}



\title{The data, they are a-changin'}

\authorinfo{Paolo Missier \and Jacek Cala \and Eldarina Wijaya}
           {School of Computing Science \\ Newcastle University}
           {\{firstname.lastname\}@ncl.ac.uk}

\maketitle

\begin{abstract}
The cost of deriving actionable knowledge from large datasets has been decreasing thanks to a convergence of positive factors:
low cost data generation, inexpensively scalable storage and processing infrastructure (cloud), software frameworks and tools for massively distributed data processing, and parallelisable data analytics algorithms. 
One observation that is often overlooked, however, is that each of these elements is not immutable, rather they all evolve over time.
This suggests that the value of such derivative knowledge may decay over time, unless it is preserved by reacting to those changes. Our broad research goal is to develop models, methods, and tools for selectively reacting to changes by balancing costs and benefits, i.e. through complete or partial re-computation of some of the underlying processes.
In this paper we present an initial model for reasoning about change and re-computations, and show how analysis of detailed provenance of derived knowledge informs re-computation decisions.
We illustrate the main ideas through a real-world case study in genomics, namely on the interpretation of human variants in support of genetic diagnosis.
\end{abstract}




\section{Introduction}  \label{sec:intro}

One simple but rarely addressed observation in Data Science is that many of the large datasets used to derive knowledge, i.e. through ``Big Data'' computations, evolve over time.  This causes problems as changes in the datasets invalidate some of the insight derived from them.  
This is true for instance for the large class of predictive models generated using statistical inference (i.e. machine learning algorithms), whose accuracy when used in the wild tends to decay as the assumptions embodied by the data used for training are no longer valid.
This problem is also relevant in data-intensive science, where experimental results often come from computational pipelines or simulations that rely on observational data. In these settings, not only the underlying data, but also the algorithms, external reference data sources used in the analysis, as well as other dependencies evolve continuously.
These changes may represent both a threat, i.e. when a stale model is used to make decisions, and an opportunity, namely to upgrade derived knowledge by performing the analysis again.
When the processes are computationally expensive and the available budget for re-doing old work is limited, it is important to be able to determine when re-computation, partial or complete, of the underlying analytic tasks in reaction to changes is beneficial.

The potential for exploiting provenance records for partial re-computation has been studied before, in the specific context of database operations. 
In the Panda system \cite{Ikeda2011,Ikeda2010}, for instance, one can determine precisely the fragment of a data-intensive program that needs to be re-executed in order to \textit{refresh} stale results. 
However, this requires the assumption that very granular data provenance can be collected for database operations, and that the semantics of these operations is well understood.
 
In contrast, in this paper we take a broader view and consider a scenario where (i) the computation involves any program $P$ that has dependencies on external data resources,  (ii) the program structure and details of its execution may be only partially observable (coarse vs fine-grained provenance), and (iii) the program may have been executed many times over many inputs, producing a (large) history $H$ of past computations and results.

We note that changes in the content of the external resources may invalidate some, but not all, of the results in $H$.
Furthermore, as noted in the Panda system, when attempting to refresh the results that are affected, it may be possible to re-compute $P$ only partially. 
In this paper we show how provenance records from past computations, of varying granularity, can be used to select the precise subset of $H$ that becomes invalid when the content of external resources changes  (\textit{re-comp scope}).
We also show how the starting point for a partial re-computation of $P$ can be pinpointed.

Our specific contributions are as follows: (i) a formalisation of a re-computation framework under our assumptions, (ii) a discussion of the role of provenance and of how granular provenance translates into efficient re-computation through precise selection of the re-comp scope, and (iii) an illustration of the framework in action on a real-world process of analysis of human genetic variants.

This initial investigation is part of a larger project, \recomp, which aims to offer models for estimating the impact of changes in input and external data on the outcome of a program, in order to prioritise re-computation over the affected population vis-\`a-vis a limited budget.

\paragraph{Related Work.} As mentioned, the Panda prototype system \cite{Ikeda2011} aims at collecting and exploiting provenance to enable data refresh, by selecting the fragments of a data-intensive workflow that must be re-executed.
The focus here is on white box computations which involve database operations, which are documented using perfect and granular provenance records, and we have already commented on the broader and less rigid scope of our work.
A formal definition of correctness and minimality of a provenance trace with respect to a data-oriented workflow is proposed by members of the same group \cite{Ikeda2013}, leading to a notion of \textit{logical provenance}.
Although this may become a potentially useful building block for a future version of this work, it completely ignores the PROV data model \cite{w3c-prov-dm} which, instead, we regard as a practical foundation to enable interoperability of any provenance-based re-computation framework.

A similar perspective to Panda is taken in the Archived Metadata and Provenance Manager (AM\&PM) \cite{Gao2012a}, with a focus on database provenance and where the main evolving element is not the data but the database schema.
Accordingly, the provenance of schema evolution is captured and can be queried, along with the provenance of the data in the current and past versions of a database. 
Interestingly, using a schema evolution language (Prism \cite{Curino:2008:GDS:1453856.1453939}) leads to a formal definition of what in this paper we  call a $\mathit{diff}$ function, aimed at quantifying the difference between two schemas.  That research is vaguely related to our work, which does not specifically address database operations placing schema evolution out of scope.


Finally, as an infrastructure mechanism to enable selective recomputation, the \textit{strong links} approach of~\cite{Koop2010} is relevant in this context.

\begin{figure*}
\centering
\begin{minipage}[b]{0.45\linewidth}
\includegraphics[height=4.5cm]{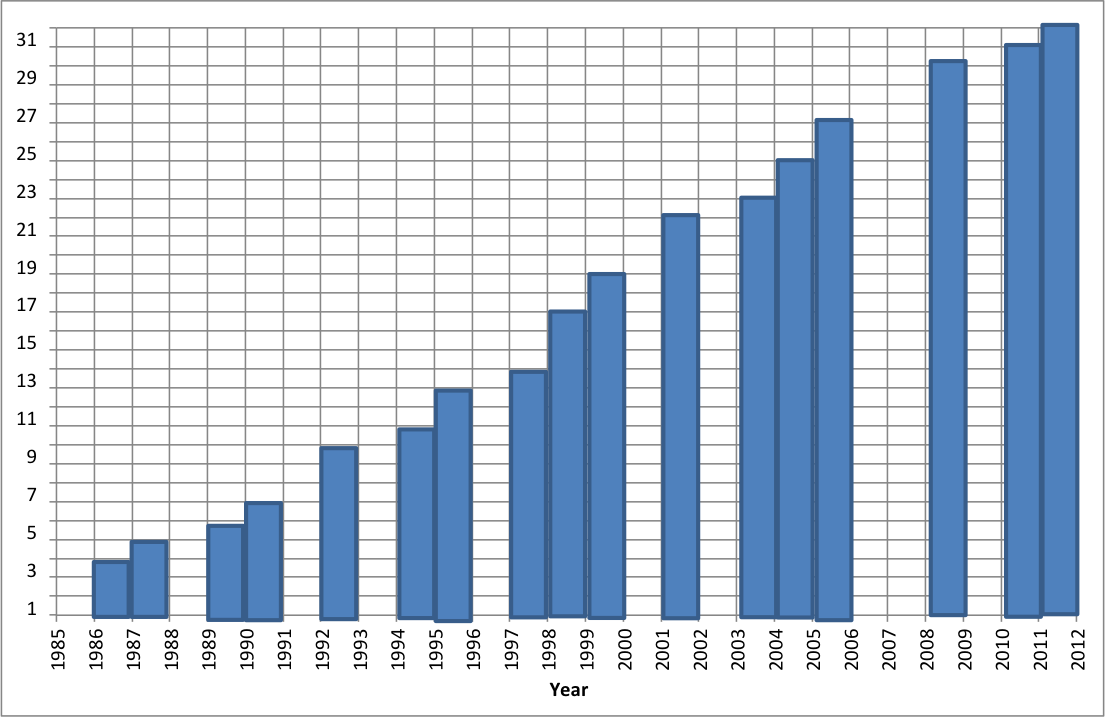}
\end{minipage}
\quad
\begin{minipage}[b]{0.45\linewidth}
\includegraphics[height=4.5cm]{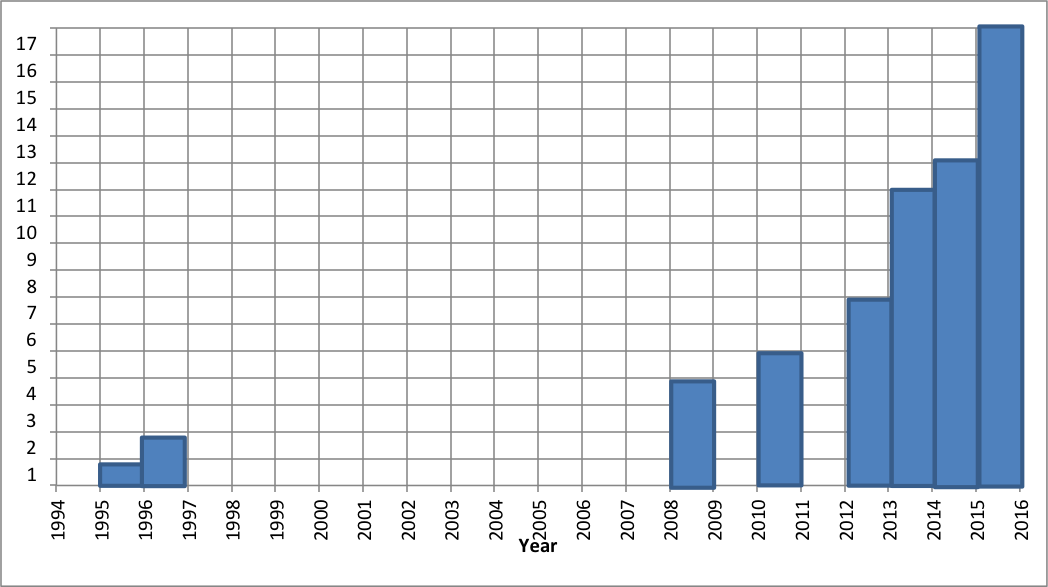}
\end{minipage}
\caption{Increase in the count of genes (left) and variants (right) over time, related to diseases affecting a cohort of patients at the IGM, Newcastle.}
\label{fig:omim-clinvar}
\end{figure*}

\paragraph{Example: analysis of human genetic variants.}
To illustrate the problem addressed in the rest of this paper, we consider a use case consisting of a process called Simple Variant Interpretation (SVI). SVI is a tool we have developed as part of the \textit{Cloud-eGenome} project \cite{Missier2015}, aimed at providing a simple interpretation of human variants to facilitate clinical diagnosis of genetic diseases. 
Here a \textit{variant}  is a single nucleotide mutation that occurs on a gene. Variants are identified by processing a patient's genome (or their \textit{exome}, a small portion of the genome) using a sequence of algorithmic steps that, essentially, compare it to a reference genome.
SVI takes a set of variants found in the patient's exome (about 25,000) and a set of terms that describe the patient's \textit{phenotype}, which indicates the patient's \textit{disease hypothesis} (presumed disorder). It selects a small subset of the variants which are relevant for the phenotype, and associates a degree of estimated deleteriousness to each of them. 
To do this it uses knowledge from external data resources, namely the ClinVar\footnote{\url{http://www.ncbi.nlm.nih.gov/clinvar/}} and OMIM Gene Map\footnote{\url{http://www.ncbi.nlm.nih.gov/omim}} databases, described in more detail later.

\begin{figure*}
\centering
\begin{minipage}[b]{0.45\linewidth}
\includegraphics[height=4.5cm]{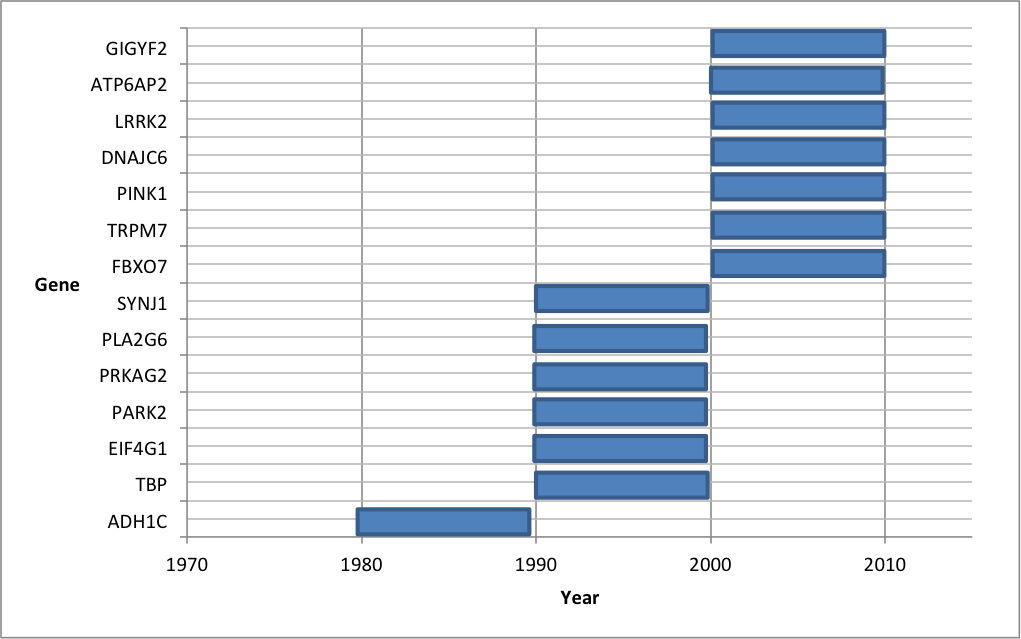}
\end{minipage}
\quad
\begin{minipage}[b]{0.45\linewidth}
\includegraphics[height=4.5cm]{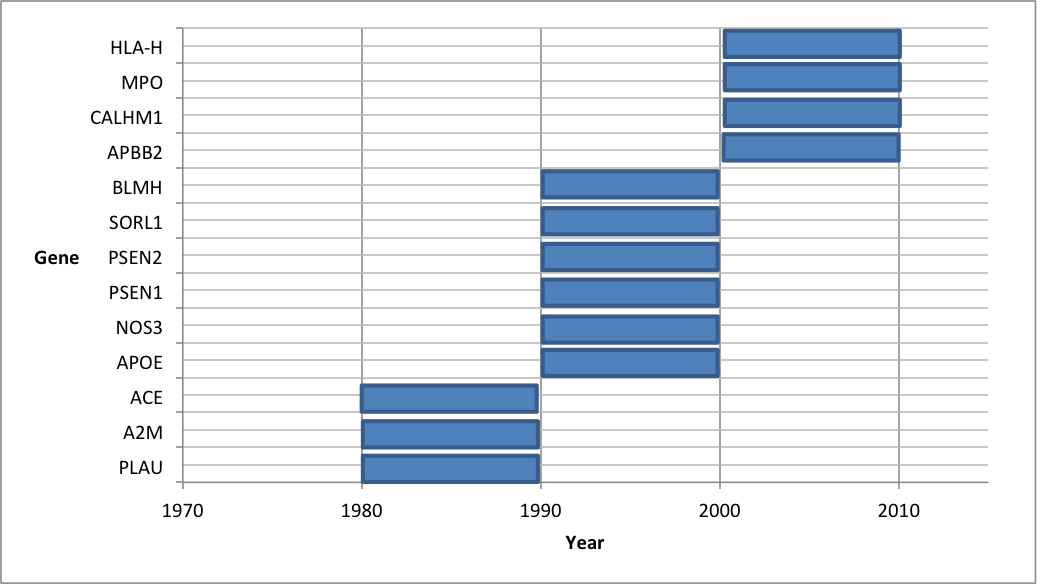}
\end{minipage}
\caption{Progressive increase in the count of genes known to be involved in Parkinson's and Alzheimer's over time, in the OMIM Gene Map database.}
\label{fig:omim-clinvar-evolution}
\end{figure*}

In some cases, the presence of deleterious variants represents conclusive evidence in support of the disease hypothesis. 
More often, however, the diagnosis is not conclusive due to missing information about the variants, or to lack of knowledge about the association between the hypothesis and the variants. 
Thus, the diagnostics conclusions that can be drawn from the data are very much dependent on the content of these reference databases which encode the current genetic knowledge. 
As this knowledge evolves and these resources are updated, there are opportunities to revisit past inconclusive diagnoses, and thus to consider re-computation of the associated analysis.
To appreciate the effect of changes in the reference knowledge, in Fig.~\ref{fig:omim-clinvar} we show how new additions to the OMIM and ClinVar databases would have affected the ability to carry out a conclusive diagnosis on a cohort of patients seen at the Institute of Genetic Medicine (IGM) in Newcastle.
The charts show the number of genes and variants within a gene, respectively, known to researchers and which would have been relevant for those patients.
The charts in Fig. \ref{fig:omim-clinvar-evolution} provide a similar view of the evolution over time of the genes known to be implicated in specific diseases, namely Parkinson's and Alzheimer's.

The \recomp\ problem in this use case involves (i) selecting the cases that are likely to benefit from re-computation, (ii) deciding whether complete or partial re-computation is required, and (iii) actually reproducing the original process, possibly requiring a new deployment.
 
\section{Re-computation framework}
To frame the re-computation decision problem in simple, abstract terms we present the set of basic elements that \recomp\ is built upon.

\paragraph{Computation.}
Consider program $P$ executing on a set $\x = \{x_1 \ldots x_n\}$ of inputs and producing outputs $\y = \{ y_1 \dots y_m\}$, which also makes use of external data resources, or \textit{data dependencies} $\D = \{D_1 \ldots D_m\}$ where each $D_i$ is a dataset, $D_i = \{ d_{i1}, d_{i2} \dots \}$.
We also associate a version $v$ to each execution. This indicates a timestamp and uniquely identifies one execution of $P$, denoted by:
\begin{equation}
 \y^{v} = P^v(\x^v | \D^v)
 \label{eq:exec-v}
\end{equation}

\paragraph{Transparency.}
The level of detail available in observing a computation of $P$ (the \textit{transparency} of $P$) plays an important role in making re-computation decisions.
We consider two aspects of transparency, namely (a) details on the internal structure of $P$, and (b) details on which subset of each $D_i$ are used.
At one end of the ``transparency spectrum'', no details are available for either (a) or (b): $P$ is a black box providing no details about its internal structure, and all we know about $D_i$ are coarse-grain statements like ``ClinVar was used at some point''. 
On the opposite end of the spectrum, $P$ is a \textit{white-box}, described for instance $P$ by function composition $P \equiv P_r \circ \ldots \circ P_1$. 
At the same time, we also understand the semantics of each subprocess $P_j$ and know the subset of $D_i$ that was used by a $P_j$.

\paragraph{Provenance.}
The provenance  of an output $\y$, denoted $\prov(\y)$, is a PROV document that describes the derivation of $\y$ from $\x$ through $P$ using elements of $\D$. 
The granularity of PROV assertions may differ depending on the transparency of $P$.
In the most granular case, when $P$ is a white box we can for instance express the usage of any single element $d_{ij} \in D_i \in \D$ by an activity $P_j$, i.e. using statements of the form:\footnote{PROV also allows to express that the $d_{ij}$ are members of a \textit{collection} $\D_i$.}
\begin{equation}
\small \texttt{used}(P_j, d_{ij}, [\texttt{prov:role} = \texttt{'dep'}])  \label{eq:pattern-dep} 
\end{equation}
where the role indicates that $d_{ij}$ is a dependency.
Similarly, for inputs $x_i$ (or intermediate values) we can write:
\begin{equation}
\small \texttt{used}(P_j, x_i, [\texttt{prov:role} = \texttt{'input'}])  \label{eq:pattern-in} 
\end{equation}
At the other extreme, in a completely black box scenario, the assertions will be of the form:
\begin{small}
\begin{align}
\texttt{used}(P, \D, [\texttt{prov:role} = \texttt{'dep'}])  \label{eq:pattern-black-dep} \text{ (use of dependency) } \\
\texttt{used}(P, \x, [\texttt{prov:role} = \texttt{'input'}])  \label{eq:pattern-black-in} \text{ (use of input) } 
\end{align}
\end{small}

In addition to producing $\prov(\y)$, each computation of form (\ref{eq:exec-v}) also generates \textit{history record} $h$:
\begin{equation}
h(\y,v) = \langle P^v, \D^v, \x^v, \prov(\y^v), \mathit{cost}(\y^v)\rangle 
\label{eq:CH}
\end{equation}
where it is expected that $\prov(\y^v)$ contains statements that make references to $P^v$, $\x^v$, and $\D^v$.
Over time, statements of the form (\ref{eq:CH}) form a \textit{History database} $\CH$.
Note that we also record the $\mathit{cost}(\y)$ of computing $\y$ by executing $P$ on $\x$.
Although the specific form of the cost is immaterial here, in practice it can be expressed as a monetary cost (e.g. when $P$ is executed on a public cloud), execution time, resource usage or as a combination of them.

\paragraph{Change detection.}

\recomp\ relies on the capability to detect and quantify changes between any two versions of $\x$ and $\D$, i.e. $\x^{v} \rightarrow \x^{v'}$, $\D^{v} \rightarrow \D^{v'}$.
Thus, we assume there exist three families of \textit{diff} functions that are needed to compare two versions of the elements of $\x$,  $\D$, and $\y$. 
\begin{align*}
\text{input diff: } & \{ \diffin(x_i^v, x_i^{v'})  | x^{v}_i \in \x^{v}, x^{v'}_i \in \x^{v'}    \} \\
\text{dependency diff: } & \{ \diffd(D_i^v, D_i^{v'}) | D^{v}_i \in \D^{v}, D^{v'}_i \in \D^{v'} \} \\ 
\text{output diff: } & \{ \diffo(y_i^v, y_i^{v'}) | y^v_i \in \y^v,  y^{v'}_i \in \y^{v'} \}
\end{align*}
These operate independently on each input, dependency, and output component. 
Each of these functions will have a different signature, and produce a summary of changes found in its inputs, in a format that may vary depending on the types of $\x$ and $\D$.
For instance, $\diffd(D_i^v, D_i^{v'})$ typically computes the symmetric difference $( D_i^v \setminus D_i^{v'}) \cup (D_i^{v'}\setminus D_i^v)$.
Other types of \textit{diff} functions can be defined for specific use cases. 
Note that, although changes in the structure of program $P$ are also relevant and are within the general \recomp\ framework, for simplicity in this short contribution we are going to assume that $P$ does not change.

\paragraph{Role of the $\CH$ database and of provenance.}  \label{sec:provenance}

As mentioned earlier, upon detecting changes (i.e. using the \textit{diff} functions) the first steps in making re-computation decisions include 
(i) \textit{scoping rules}, that is selecting the subset $H' \subset H$ of the computations described in $H$ that are affected by these changes, and 
(ii) defining the starting point $P_s$ of a \textit{partial} re-computation of $P$, which we call the \textit{starting component} of $P$. 
This is the component of $P$ mentioned in the earliest usage of a changed dataset (input or dependency), and it is not necessarily the same as the start of the whole of $P$.
Note that partial re-computation is only possible if the input to $P_s$ is available, i.e. not only should the input be explicitly mentioned in $\prov(\y)$, but it must also have been cached in a data store.

In a white box scenario where $P$'s executions are fully observable, both steps can be addressed by querying the provenance documents in $\CH$.
We distinguish the case of a change in inputs $\x$ from the case of a change in a dependency $D_i \in \D$.
These correspond to the two patterns (\ref{eq:pattern-in}) and (\ref{eq:pattern-dep}) above.
Specifically, if the change $x_i^v \rightarrow x_i^{v'}$ involves any of the inputs $x_i \in \x$, the scope $H'$ is simply the set of records $h$ in which $x_i^v$ is used as input, i.e. all
$h(\y, v)$ such that $\prov(\y^v)$ includes the pattern of form (\ref{eq:pattern-in}).

Regarding dependency change $D_i^v \rightarrow D_i^{v'}$, the affected records are those where the computation involved elements in $\diffd(D_i^v, D_i^{v'})$. 
These are the $h(\y,v)$ such that: (i) $\prov(\y^v)$ includes the pattern of form (\ref{eq:pattern-dep}) involving data element $d_{ij}$, and (ii) 
$d_{ij} \in \diffd(D_i^v, D_i^{v'})$. 

Next, within the scope determined as above, we need to determine the starting component $P_s$ of each $P$.
The provenance patterns just mentioned, (\ref{eq:pattern-in}) and (\ref{eq:pattern-dep}), provide the answer, namely the starting component is the activity $P_j$ that appears in the \textit{earliest} occurrence of a usage statement involving a changed input or dependency.

Finally, note that in a black box scenario, with either limited visibility of process structure and/or of data input granularity, the scoping rules cannot be used, i.e. the default scope is the whole of $H$, and total (as opposed to partial) re-computation of $P$ is required.

\begin{figure*}
  \centering
  \includegraphics[width=.8\linewidth]{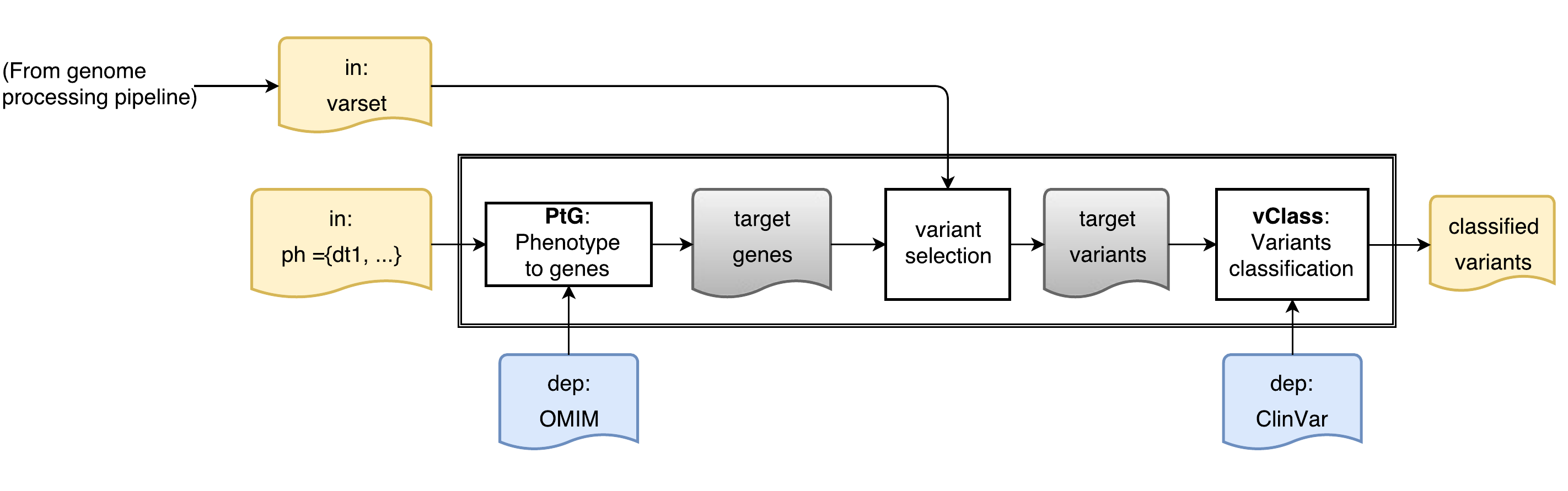}
  \caption{\textit{White box} SVI, with inputs $\x = [\mathit{varset}, \mathit{ph}]$ and data dependencies $\D = [\text{OMIM}, \text{ClinVar}]$}
  \label{fig:svi-pipeline}
\end{figure*}

\section{Detailed use case: SVI re-computation}

We now illustrate the framework in use on our SVI case study.
One execution of SVI, illustrated in Fig. \ref{fig:svi-pipeline}, is carried out for each patient whose diagnosis we want to confirm.
SVI is an example of process $P$ with inputs:
\[\x = [\mathit{varset}, \mathit{ph}]\]
where  $\mathit{varset}$ is the set of variants associated with the patient, and $\mathit{ph} = \{ \dt_1, \dt_2, \dots \}$ is the phenotype expressed using \textit{disease terms} $\dt_i$ from the OMIM vocabulary, for example \emph{Alzheimer's}.\footnote{Terms from HPO, the Human Phenotype Ontology, are also used in practice.}
SVI is a classifier that associates a class label to each input variant depending on their estimated deleteriousness, using a simple ``traffic light'' notation, i.e.:
\[\y = \{ (v, \mathit{class}) | v \in  \mathit{varset}, \mathit{class} \in \{ \textsf{red}, \textsf{amber}, \textsf{green}\}\} \]
The \textsf{amber} variants are typically the majority, indicating uncertain diagnosis, while one or two \textsf{red} (pathogenic) variants are sufficient to complete the diagnosis.

SVI's data dependencies $\D$ consist of the two reference databases, OMIM and Clinvar, each subject to periodic revisions and denoted 
$\D^v = [\OM{v}, \CV{v}]$.
OMIM maintains a vocabulary $\DT$ of standard terms used to denote human disorders.
To each of these terms, OMIM associates a set of genes that are known to be broadly involved in the disease.
We denote the mapping from $\dt \in \DT$ to genes via $\OM{v}$ as $\genes(\dt, \OM{v})$.
ClinVar contains catalogue $V$ of single-nucleotide human variants, and it is used to try and determine the clinical significance of the genetic mutations in patients, i.e. those variants found in $\mathit{varset}$.
Specifically, Clinvar associates a status to each variant $\mathit{var} \in V$, denoted
$\varst(\mathit{var}, \CV{v}) \in \{ \textsf{unknown}, \textsf{benign}, \textsf{pathogenic} \}$.


SVI uses versions $\OM{v}$ and $\CV{v}$ of OMIM and ClinVar to investigate a patient's disease, as shown in Fig. \ref{fig:svi-pipeline}.  
Firstly, the terms in $\mathit{ph}$ are used to determine the set of \textit{target genes} that are relevant for the disease hypothesis. 
These are defined as the union of all the genes in $genes(\dt, \OM{v}) $ for each disease term $\dt \in  \mathit{ph}$.
For example, if the patient's phenotype is \emph{Alzheimer's}, then genes like \texttt{PSEN2} and \texttt{PLAU} are in scope, but for instances a gene that is  known to be implicated in \emph{Parkinson's disease}, such as \texttt{PARK2}, will not be considered in the investigation.

Secondly, a subset of the variants in $\mathit{varset}$ is selected, according to standard filtering rules;
variant $\mathit{var} \in \mathit{varset}$  is selected if it is located on the \textit{target genes}.
And, finally, the variants are classified as red, amber, or green depending on $\varst(\mathit{var}, \CV{v})$ (variants that are not catalogued at all in ClinVar become amber).

To illustrate the process consider two patients from the IGM cohort mentioned briefly in the introduction.
Patient 1 is diagnosed with Alzheimer's, while Patient 2 is presumably affected by Parkinson's. 
Since the ’90s, two genes have been known to be loosely implicated in these diseases, \texttt{PSEN2} and \texttt{PARK2}, respectively:

\begin{align*}
\texttt{PSEN2} \in \genes(\texttt{Alzheimer's}, \OM{1995}), \\
\texttt{PARK} \in \genes(\texttt{Parkinson's}, \OM{1995}) 
\end{align*}

However, it was not until 2015 that two specific variants situated on those genes, at position 227083249 and 161807855, respectively have been studied and added to ClinVar. 
Thus, until 2014 we had 
\begin{align*}
\varst(227083249, \CV{2014}) = \textsf{amber}, \\
\varst(161807855, \CV{2014}) = \textsf{amber}
\end{align*}
because neither variants were known to ClinVar.

\paragraph{Diff functions.}

For OMIM, $\diffOM(\OM{v}, \OM{v'})$ returns the set of terms $t \in T$ for which the mapping to genes has changed:
\begin{align*}
 \diffOM&(\OM{v}, \OM{v'}) = \\
& \{ t \in \DT | genes(t, \OM{v}) \neq genes(t, \OM{v'}) \} 
\end{align*}
\sloppy while $\diffCV(\CV{v}, \CV{v'})$ 
returns set of variants $\mathit{var} \in V$ with changed status, as well as new variants, or removed variants:
\begin{small}
\begin{align*}
\diffCV&(\CV{v}, \CV{v'}) = \\
&\{ \mathit{var} \in V | \varst(\mathit{var}, \CV{v}) \neq \varst(\mathit{var}, \CV{v'}) \} \\
& \cup \CV{v'} \setminus \CV{v} \cup \CV{v} \setminus \CV{v'}
\end{align*}
\end{small}

\paragraph{Use of provenance.}
The provenance from each SVI tool execution is recorded in the $\CH$ database.
Following the white box approach of Sec.~\ref{sec:provenance}, the relevant PROV assertions generated from an execution of SVI, with block names as in Fig.~\ref{fig:svi-pipeline}, are as follows:
\begin{small}
\begin{align}
\texttt{entity}(\texttt{om}, [\texttt{prov:type} = \text{'OMIM'}, version=\texttt{'v'}]) \label{eq:omim-prov-1}  \\
\texttt{entity}(\texttt{ph},  [\texttt{prov:type} = \text{'prov:collection'}]) \\
\texttt{entity}(\texttt{cv}, [\texttt{prov:type} = \text{'CV'}, version=\texttt{'v'}]) \label{eq:omim-prov-6}  \\
\texttt{entity}(\texttt{vars},  [\texttt{prov:type} = \text{'prov:collection'}])  \\
\texttt{used}(\texttt{PtG}, \texttt{om}, [\texttt{prov:role} = \texttt{'dep'}])  \label{eq:omim-prov-2}\\
\texttt{used}(\texttt{PtG}, \texttt{ph}, [\texttt{prov:role} = \texttt{'input'}]) \label{eq:omim-prov-3}  \\
\texttt{used}(\texttt{vClass}, \texttt{cv}, [\texttt{prov:role} = \texttt{'dep'}])  \label{eq:omim-prov-4}\\
\texttt{used}(\texttt{vClass}, \texttt{vars}, [\texttt{prov:role} = \texttt{'input'}]) \label{eq:omim-prov-5}
\end{align}
\end{small}
Note that the \texttt{used} assertions are of the form (\ref{eq:pattern-dep}) and (\ref{eq:pattern-in}), respectively.
These provenance statements can be used to define scoping rules and starting components, as follows.

\paragraph{Re-comp scope due to OMIM changes.} The executions $h$ in the re-comp scope following change $\OM{v} \rightarrow \OM{v'}$ include those where phenotype $\mathit{ph}$ includes terms in $\diffOM(\OM{v}, \OM{v'})$, i.e., those with changes to their gene mappings.
{The phenotype is found in ({\ref{eq:omim-prov-3}}),} while the version of OMIM for computing diff is found using (\ref{eq:omim-prov-2}).
As (\ref{eq:omim-prov-2}) contains the earliest mention of \texttt{om}, \texttt{PtG} is also the starting component for re-computation.

\paragraph{Re-comp scope due to ClinVar changes.} Similarly, following change $\CV{v} \rightarrow \CV{v'}$, the executions in scope are those that include selected variants on target genes and which appear in 
$\diffCV(\CV{v}, \CV{v'})$.
Using the provenance fragment above, the selected variants are found in (\ref{eq:omim-prov-5}), and the version of CV for computing diff is found using (\ref{eq:omim-prov-4}). In this case, \texttt{vClass} is the starting component for re-computation following a change in ClinVar.

\paragraph{Example, continued.}
Continuing with our earlier two-patients example, consider again variants  227083249 and 161807855.
Because they are both located on genes that have been known to OMIM for many years, these variants are selected as candidates for testing against ClinVar. 
As mentioned, until 2014 they were both classified as \textsf{'amber'}.
Having been added to ClinVar in 2015, however, they both appear in the latest diff between the 2014 and 2015 versions of ClinVar:
\[ \{ \texttt{227083249}, \texttt{161807855} \} \subset \diffCV(\CV{2014}, \CV{2015}) \]
According to the scoping rule above, the re-comp scope due to these additions includes the executions of $h$ where the provenance mentions 227083249 and 161807855, which include patients 1 and 2 (possibly along with many other patients, but none of those for which these variants are not relevant).
As 227083249 is catalogued as ``probably pathogenic, uncertain significance'', the diagnosis for patient 1 is still inconclusive. For Patient 2, on the other hand, we can rule out variant 161807855 as a cause of their disease, as this variant is now known to be benign.

\section{Conclusions}
\textit{Knowledge assets} derived from data analytics computations may decay and become obsolete as the datasets or the content of reference data resources used to produce it change over time. 
While this suggests that re-computation of such knowledge assets may be needed, deciding precisely which of them should be re-computed is not a trivial problem;
it requires meta-knowledge about their dependencies on the inputs and on the reference datasets.

In this paper we have discussed the role of provenance in providing such meta-knowledge, in a way that can be used to inform re-computation decisions.
We have presented a simple reference framework in which data is versioned and functions are available to compute the differences between any two versions. We have clarified how fine-grained and coarse-provenance can be used to assess the impact of such differences on a history of past computations, with different precision, suggesting which past computations should be performed anew.
 We have illustrated these ideas through a detailed example, concerning the automated classification of human variants for clinical diagnosis.

\acks
This work is funded in part by EPSRC grant no. EP/N01426X/1 in the UK.





\end{document}